# Quantum dynamics in view of Einstein's theory of Brownian motion

S. ABE[1] (*) and A. K. RAJAGOPAL[2]

[1] *Institute of Physics, University of Tsukuba, Ibaraki 305-8571, Japan*

[2] *Naval Research Laboratory, Washington, DC 20375-5320, USA*

**Abstract.** - A quantum-mechanical version of Einstein's 1905 theory of Brownian motion is presented. Starting from the Hamiltonian dynamics of an isolated composite of objective and environmental systems, subdynamics for the objective system is derived in the spirit of Einstein. The resulting master equation is found to have the Lindblad structure.

Einstein, in his landmark paper [1] on classical Brownian motion, has developed for the first time the basis for the diffusion equation. His way of deriving the equation is well explicated in Ref. [2]. This theory consists of two independent distributions, one for finding a Brownian particle in space-time $(x, t)$ and the other for spatial displacement $\Delta$ of the particle in a single discrete time step $\tau$. Let us denote them to be $f(x, t)$ and $\phi(\Delta)$, respectively. The essence of Einstein's theory is in expressing temporally evolved $f(x, t + \tau)$ in terms of the average of the spatially displaced distribution $f(x + \Delta, t)$ over $\phi(\Delta)$:

$$f(x, t + \tau) = \int d\Delta \, f(x + \Delta, t) \phi(\Delta). \qquad (1)$$

Assuming analyticity of $f(x, t)$ in both space and time and evenness of $\phi(\Delta)$, and working to the leading order on both sides of Eq. (1), Einstein has obtained by-now the standard diffusion equation: $\partial f / \partial t = D \partial^2 f / \partial x^2$, where the diffusion constant is given by $D = (1/2\tau) \int d\Delta \, \Delta^2 \phi(\Delta)$. By relaxing the analyticity assumption in space, Abe and Thurner [3] have recently generalized the above theory to derive a fractional differential equation describing anomalous diffusion.

The physical basis behind this is to express the effect of the environment causing evolution as due to the distribution of displacement. The purpose of this work is to implement this idea to develop a quantum-mechanical approach to a composite of the objective and environmental systems. This is accomplished by Hamiltonian evolution of the total system described by a density matrix and reduction of it to subdynamics of the objective system. In this viewpoint, Eq. (1) is seen to govern subdynamics, which naturally leads to physical clarification of $\phi(\Delta)$ in the quantum-mechanical regime.

Consider an isolated composite of two subsystems, *A* and *B*, where *A* is the objective and *B* is the environment. The density matrix of the total system evolves in time as

$$\hat{\rho}(t+\tau) = \hat{U}(\tau)\,\hat{\rho}(t)\,\hat{U}^{\dagger}(\tau), \tag{2}$$

where $\hat{U}(\tau)$ is the unitary operator, $\hat{U}(\tau) = \exp(-i\tau\hat{H})$, with the total Hamiltonian

$$\hat{H} = \hat{H}_A + \hat{H}_B + \hat{H}_{\text{int}}. \tag{3}$$

The interaction represented by $\hat{H}_{\text{int}}$ is required to be weak. Here and hereafter, the Planck constant ($h/2\pi$) is set equal to unity. The reduced density matrix of the subsystem *A* is given by the partial trace over *B*:

$$\hat{\rho}_A(t+\tau) = \text{Tr}_B\left[\hat{U}(\tau)\hat{\rho}(t)\hat{U}^\dagger(\tau)\right], \tag{4}$$

which becomes

$$\hat{\rho}_A(t+\tau) = e^{-i\tau\hat{H}_A} \text{Tr}_B\left[\hat{V}(\tau)\hat{\rho}(t)\hat{V}^\dagger(\tau)\right] e^{i\tau\hat{H}_A}, \tag{5}$$

where

$$\hat{V}(\tau) = \text{T}\exp\left(-i\tau\hat{H}_B - i\int_0^\tau ds\, e^{i\tau\hat{H}_A}\,\hat{H}_{\text{int}}\, e^{-i\tau\hat{H}_A}\right) \tag{6}$$

with the chronological symbol T for the time-ordered product. Taking the trace over *B* by using a certain discrete basis $\{|e_k(B)\rangle\}_k$, we obtain

$$\hat{\rho}_A(t+\tau) = e^{-i\tau\hat{H}_A} \sum_k \langle e_k(B)|\hat{V}(\tau)\hat{\rho}(t)\hat{V}^\dagger(\tau)|e_k(B)\rangle\, e^{i\tau\hat{H}_A}. \tag{7}$$

Here, we make two observations in the spirit of Einstein in order to derive the corresponding quantum theory of diffusion. The environment *B* is composed of fast degrees of freedom while leaving *A* intact, which is achieved by an adiabatic scheme. Using the orthonormal basis $\{|\Delta(B)\rangle\}_\Delta$, which diagonalizes the state of *B*, we have

$$e^{i\tau \hat{H}_A} \hat{\rho}_A(t+\tau) e^{-i\tau \hat{H}_A}$$
$$= \sum_k \sum_{\Delta, \Delta'} \langle e_k(B)| \hat{V}(\tau) |\Delta(B)\rangle \langle \Delta(B)| \hat{\rho}(t) |\Delta'(B)\rangle \langle \Delta'(B)| \hat{V}^\dagger(\tau) |e_k(B)\rangle. \quad (8)$$

The basis $\{|\Delta(B)\rangle\}_\Delta$, which is different from an arbitrary one $\{|e_k(B)\rangle\}_k$, is employed in order to take account of the fast degrees of freedom of *B*. Then, the adiabatic scheme as well as the assumptions of weakness of the interaction and entanglement between *A* and *B* allows us to write the matrix elements in the *B*-space as follows:

$$\langle \Delta(B)| \hat{\rho}(t) |\Delta'(B)\rangle = \hat{\rho}_A(t) \varphi(\Delta) \delta_{\Delta, \Delta'}, \quad (9)$$

where $\sum_\Delta \varphi(\Delta) \cong \int d\Delta \, \varphi(\Delta) = 1$, provided the basis is assumed to be quasi-continuous for a large environment. The use of the same symbol $\Delta$ as that in Eq. (1) may not lead to confusion. Defining

$$\hat{M}_k(\Delta; \tau) \equiv \langle e_k(B)| \hat{V}(\tau) |\Delta(B)\rangle, \quad (10)$$

we arrive at the Kraus-type representation [4]

$$e^{i\tau\hat{H}_A}\,\hat{\rho}_A(t+\tau)\,e^{-i\tau\hat{H}_A} = \sum_k \int d\Delta\,\hat{M}_k(\Delta;\tau)\,\hat{\rho}_A(t)\,\hat{M}_k^\dagger(\Delta;\tau)\,\varphi(\Delta). \qquad (11)$$

This is the quantum-mechanical counterpart of Eq. (1) and accordingly $\varphi(\Delta)$ corresponds to $\phi(\Delta)$ appearing therein. From Eq. (9), the physical significance of the distribution $\varphi(\Delta)$ is now clear to be due to the adiabatic scheme of treating fast dynamics of the environment in relation to that of the objective system and the approximate product structure of the total density matrix. We note the important positive semi-definiteness of the right-hand side of Eq. (11), which is essential for subdynamics to be a completely positive quantum operation.

Performing the trace operations of both sides of Eq. (11), we have the trace-preserving condition

$$\hat{I}_A = \sum_k \int d\Delta\,\hat{M}_k^\dagger(\Delta;\tau)\,\hat{M}_k(\Delta;\tau)\,\varphi(\Delta), \qquad (12)$$

where $\hat{I}_A$ is the identity operator in the $A$-space. This implies that

$\left\{ \hat{M}_k^\dagger(\Delta; \tau) \, \hat{M}_k(\Delta; \tau) \, \varphi(\Delta) \right\}_{k, \Delta}$ is a positive operator-valued measure.

Taking the anticommutator of Eq. (12) with $\hat{\rho}_A(t)/2$ and subtracting it from Eq. (11), we have

$$e^{i\tau \hat{H}_A} \, \hat{\rho}_A(t+\tau) \, e^{-i\tau \hat{H}_A} - \hat{\rho}_A(t) = \sum_k \int d\Delta \, \varphi(\Delta) \, L_k(\Delta; \tau)[\hat{\rho}_A(t)], \tag{13}$$

where $L_k(\Delta; \tau)$ is the superoperator defined by

$$L_k(\Delta; \tau)[\hat{\rho}_A] \equiv \hat{M}_k(\Delta; \tau) \, \hat{\rho}_A \, \hat{M}_k^\dagger(\Delta; \tau)$$
$$- \frac{1}{2} \hat{\rho}_A \, \hat{M}_k^\dagger(\Delta; \tau) \, \hat{M}_k(\Delta; \tau) - \frac{1}{2} \hat{M}_k^\dagger(\Delta; \tau) \, \hat{M}_k(\Delta; \tau) \, \hat{\rho}_A \, . \tag{14}$$

Expanding the left-hand side of Eq. (13) with respect to $\tau$ and keeping the leading order contributions, we have

$$\frac{\partial \hat{\rho}_A(t)}{\partial t} + i\left[\hat{H}_A, \hat{\rho}_A(t)\right] = \frac{1}{\tau} \sum_k \int d\Delta \, \varphi(\Delta) \, L_k(\Delta; \tau)[\hat{\rho}_A(t)]. \tag{15}$$

Let us assume that the superoperator is analytic in $\Delta$ so that

$$\hat{M}_k(\Delta; \tau) = \sum_{n=0}^{\infty} \Delta^n \, \hat{\Lambda}_k^{(n)}(\tau), \tag{16}$$

and $\varphi(\Delta)$ has the moments of all orders, i.e., $\langle \Delta^n \rangle = \int d\Delta\, \Delta^n\, \varphi(\Delta) < \infty$ $(n = 0, 1, 2, \cdots)$.

Then, Eq. (15) becomes

$$\frac{\partial \hat{\rho}_A(t)}{\partial t} + i\left[\hat{H}_A, \hat{\rho}_A(t)\right] = \sum_k \sum_{m,n=0}^{\infty} D_{m+n}\, K_k^{(m,n)}(\tau)\,[\hat{\rho}_A(t)], \tag{17}$$

where $K_k^{(m,n)}(\tau)$ is the superoperator given by

$$K_k^{(m,n)}(\tau)[\hat{\rho}_A] = \hat{\Lambda}_k^{(m)}(\tau)\, \hat{\rho}_A\, \hat{\Lambda}_k^{(n)\dagger}(\tau) - \frac{1}{2}\hat{\rho}_A\, \hat{\Lambda}_k^{(n)\dagger}(\tau)\, \hat{\Lambda}_k^{(m)}(\tau) - \frac{1}{2}\hat{\Lambda}_k^{(n)\dagger}(\tau)\, \hat{\Lambda}_k^{(m)}(\tau)\, \hat{\rho}_A, \tag{18}$$

and $D_{m+n}$ is the generalized diffusion constant

$$D_{m+n} = \frac{\langle \Delta^{m+n} \rangle}{\tau}. \tag{19}$$

$K_k^{(m,n)}(\tau)$ can also be expanded with respect to $\tau$, but it is left as it is, here. Thus, we see that the resulting master equation governing subdynamics of the objective system has the Lindblad structure, which maintains the Markovian nature as well as the properties to be satisfied by a density matrix (i.e., Hermitian, positive and unit trace).

In conclusion, we have developed the quantum mechanical version of Einstein's 1905 theory of Brownian motion and have derived the master equation for the subsystem, starting from the Hamiltonian. This master equation is shown to have the Lindblad structure. In this way, we have clarified the physical origin of the analog of the distribution of displacement in Einstein's theory in terms of a microscopic quantum-mechanical description of the environment.

***

S. A. thanks Naval Research Laboratory for hospitality extended to him, which enabled this work to be completed. A. K. R. acknowledges partial support of the Office of Naval Research.

REFERENCES

[1]   EINSTEIN A., *Ann. Phys. (Leipzig)*, **17** (1905) 549. English translation: *Investigations on the Theory of Brownian Movement* (Dover, New York) 1956.

[2]   PAIS A., *'Subtle is the Lord ...'* (Oxford University Press, Oxford) 1982.

[3]  ABE S. and THURNER S., *Physica A*, **356** (2005) 403.

[4]  KRAUS K., *Ann. Phys. (NY)*, **64** (1971) 311.

[5]  LINDBLAD G., *Commun. Math. Phys.*, **48** (1976) 119.

[6]  GORINI V., KOSSAKOWSKI A. and SUDARSHAN E. C. G.,

   *J. Math. Phys.*, **17** (1976) 821.